\documentclass[sigconf,screen]{acmart}

\usepackage{framed}
\usepackage{mdframed}
\usepackage{algorithm}
\usepackage{algorithmic}
\usepackage{multirow}
\usepackage{threeparttable}
\usepackage{xspace}
\usepackage{xcolor}
\usepackage{listings}
\usepackage{colortbl}
\usepackage{booktabs}
\usepackage[inkscapelatex=false]{svg}
\usepackage{multicol}
\usepackage{balance}
\usepackage{hyperref}
\usepackage{cleveref}

\definecolor{codegreen}{rgb}{0,0.6,0}
\definecolor{codegray}{rgb}{0.5,0.5,0.5}
\definecolor{codepurple}{rgb}{0.58,0,0.82}
\definecolor{backcolour}{rgb}{0.95,0.95,0.92}
\definecolor{light-gray}{gray}{0.95}

\lstdefinestyle{mystyle}{
  commentstyle=\color{codegreen},
  keywordstyle=\color{magenta},
  numberstyle=\tiny\color{codegray},
  stringstyle=\color{codepurple},
  basicstyle=\ttfamily\footnotesize,
  breakatwhitespace=false,         
  breaklines=true,                 
  captionpos=b,                    
  keepspaces=true,                 
  numbers=left,                    
  numbersep=-2pt, 
  columns=flexible,                  
  tabsize=4
}
\newcommand{\ie}{\textit{i}.\textit{e}.,\xspace}
\newcommand{\eg}{\textit{e}.\textit{g}.,\xspace}
\newcommand{\aka}{\textit{a}.\textit{k}.\textit{a}.,\xspace}
\newcommand{\cf}{\textit{cf}.\xspace}

\AtBeginDocument{%
  \providecommand\BibTeX{{%
    \normalfont B\kern-0.5em{\scshape i\kern-0.25em b}\kern-0.8em\TeX}}}

\begin{document}

\title[Detecting Build Dependency Errors in Incremental Builds]{Detecting Build Dependency Errors in Incremental Builds
}

\title{Detecting Build Dependency Errors in Incremental Builds}

\author{Jun Lyu}
\affiliation{%
  \institution{Nanjing University}
  \city{Nanjing}
  \country{China}
}
\email{lvjun@smail.nju.edu.cn}

\author{Shanshan Li}
\authornote{Corresponding author}
\affiliation{%
  \institution{Nanjing University}
  \city{Nanjing}
  \country{China}
}
\email{lss@nju.edu.cn}

\author{He Zhang}
\affiliation{%
  \institution{Nanjing University}
  \city{Nanjing}
  \country{China}
}
\email{hezhang@nju.edu.cn}

\author{Yang Zhang}
\affiliation{%
  \institution{Nanjing University}
  \city{Nanjing}
  \country{China}
}
\email{zhangyang2021@smail.nju.edu.cn}

\author{Guoping Rong}
\affiliation{%
  \institution{Nanjing University}
  \city{Nanjing}
  \country{China}
}
\email{ronggp@nju.edu.cn}

\author{Manuel Rigger}
\affiliation{%
  \institution{National University of Singapore}
  \city{Singapore}
  \country{Singapore}
}
\email{rigger@nus.edu.sg}

\setcopyright{acmlicensed}
\acmDOI{10.1145/3650212.3652105}
\acmYear{2024}
\copyrightyear{2024}
\acmSubmissionID{issta24main-p38-p}
\acmISBN{979-8-4007-0612-7/24/09}
\acmConference[ISSTA '24]{Proceedings of the 33rd ACM SIGSOFT International Symposium on Software Testing and Analysis}{September 16--20, 2024}{Vienna, Austria}
\acmBooktitle{Proceedings of the 33rd ACM SIGSOFT International Symposium on Software Testing and Analysis (ISSTA '24), September 16--20, 2024, Vienna, Austria}
\received{16-DEC-2023}
\received[accepted]{2024-03-02}


\begin{abstract}
Incremental and parallel builds performed by build tools such as Make are the heart of modern C/C++ software projects. Their correct and efficient execution depends on build scripts.
However, build scripts are prone to errors. The most prevalent errors are missing dependencies (MDs) and redundant dependencies (RDs). The state-of-the-art methods for detecting these errors rely on clean builds (\ie full builds of a subset of software configurations in a clean environment), which is costly and takes up to a few hours for large-scale projects.
To address these challenges, we propose a novel approach called \textsc{EChecker} to detect build dependency errors in the context of incremental builds. The core idea of \textsc{EChecker} is to automatically update actual build dependencies by inferring them from C/C++ pre-processor directives and Makefile changes from new commits, which avoids clean builds when possible. \textsc{EChecker} achieves higher efficiency than the methods that rely on clean builds while maintaining effectiveness.  
We selected 12 representative projects, with their sizes ranging from small to large, with 240 commits (20 commits for each project), based on which we evaluated the effectiveness and efficiency of \textsc{EChecker}. We compared the evaluation results with a state-of-the-art build dependency error detection tool. The evaluation shows that the F-1 score of \textsc{EChecker} improved by 0.18 over the state-of-the-art method.
\textsc{EChecker} increases the build dependency error detection efficiency by an average of 85.14 times (with a median of 16.30 times). The results demonstrate that \textsc{EChecker} can support practitioners in detecting build dependency errors efficiently.
\end{abstract}

\maketitle

\section{Introduction}

Common build systems~\cite{Licker19}, such as \textsc{GNU Make}~\cite{GNUmake20}, \textsc{SCons}~\cite{scons23}, \textsc{Ant}~\cite{Ant21}, \textsc{Maven}~\cite{maven23} and \textsc{Gradle}~\cite{gradle23}, rely on build scripts to perform builds correctly and efficiently. However, developing such build scripts is not an easy task. Although automated script generators such as \textsc{Autotools}~\cite{auto20} and \textsc{CMake}~\cite{CMake10} may help, there is still a need to manually enumerate build dependencies for custom software. For real large-scale projects, dependency enumeration is error-prone. 
In particular, in C-based projects, 52.68\% of the build errors are dependency-related errors~\cite{Seo14}.

Common build dependency errors for C-based projects are \textit{Missing Dependencies} (MDs) and \textit{Redundant Dependencies} (RDs). The root causes of both dependency errors are the differences between the dependencies declared in the build script (\aka declared build dependencies) and the dependencies used by the build system in the actual build (\aka actual build dependencies). Actual build dependencies refer to the dependencies used by the build system in the build process of the targets~\cite{Actual22}. MDs refer to the actual build dependencies that are erroneously missing from the declared build dependencies. \textsc{GNU Make}, which is one of the most widely-used build systems~\cite{McIntosh11,McIntosh14,McIntosh15}, performs incremental builds based on the dependencies declared in the build script~\cite{GNUmake20}. MDs prevent \textsc{GNU Make} from recompiling programs after they have been modified and regenerating all the targets that contain them, resulting in incorrect incremental builds~\cite{Licker19}. RDs refer to declared build dependencies which declare dependencies that are not the actual build dependencies of the target. RDs cause the build system to perform unnecessary incremental builds. In addition, RDs enforce the targets that could be executed in parallel to be executed sequentially, reducing the build efficiency~\cite{Licker19}.

The software engineering (SE) community has proposed many methods to detect build dependency errors~\cite{Zhou14,Xia14}. 
The state-of-the-art methods utilize a clean build to obtain the actual build dependencies~\cite{Licker19,Bezemer17,Sotiropoulos20,Fan20,Wu22}. Such methods often detect dependency errors using actual build dependencies together with or without declared build dependencies. Running such a tool for each commit is desirable because each commit can introduce dependency errors and it is more cost-effective to fix an error as early as possible in the development process. The current state-of-the-art methods require a clean build for each detection, making this approach costly. Clean builds can be very time-consuming for large projects (\eg \textit{OpenCV} clean builds on our machine in over an hour). It is difficult for practitioners to quickly perform build dependency error detection with state-of-the-art methods for each commit.

\begin{figure}[!t]
  \centering
  \includegraphics[scale=0.5,trim=0 430 530 10,clip]{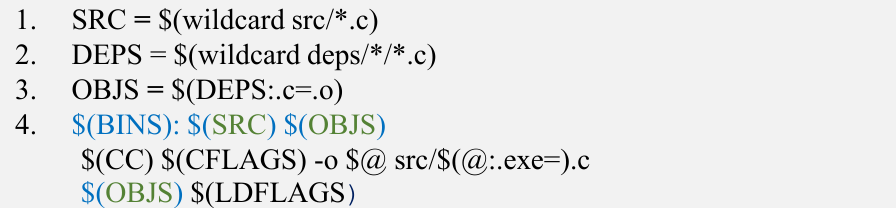}
  \caption{Dependency errors form \textit{Clib} Makefile}
  \label{FIG:errors_example}
  \vspace*{-3.0ex}
\end{figure}

Consider the following example from the \textit{clib} Makefile shown in \Cref{FIG:errors_example}. 
The build script (Makefile) declares all files with the suffix ``.c'' as prerequisites (see line 4), but no other files (\eg header files) are included. Suppose a practitioner changes a header. \textsc{GNU Make} will not regenerate any of the files containing the header. This means that the header files are MDs, which may result in incorrect builds. Changes to MDs files that should be rebuilt and relinked in an incremental build may not trigger the correct incremental build as expected.
As with other state-of-the-art approaches, Buildfs~\cite{Sotiropoulos20} performs a clean build of the project once per detection.
Even if a new commit implements only a minor modification of one of the header files, the current methods still require a clean build to detect dependency errors. The clean build is time-consuming for large-scale projects, rendering the dependency detection methods relying on it non-interactive and expensive.

To address these challenges, we propose an approach called \textsc{EChecker} to detect build dependency errors for C-based and \textsc{GNU Make}-based projects. 
The core idea behind \textsc{EChecker} is to infer actual build dependency changes based on code changes in commits. In detail, actual build dependencies can be inferred from pre-processor directives and Makefile changes based on new commits, rather than obtaining actual build dependencies by executing an entire clean build.
Our key observation is that the actual build dependencies of the target only change when the pre-processor directives~\cite{Preprocessor23} of source code (include directive, \ie \texttt{\#include <example.h>}) and the build commands in the Makefile are changed. The pre-processor directives and build commands indicate which files the compiler should include in the source file for building, \ie they indicate the actual build dependencies. 
However, this is insufficient to detect some dependency errors when dependencies only emerge during building  (\eg a header file might be automatically generated in the build, called hidden dependencies)~\cite{Licker19}.
Thus, \textsc{EChecker} monitors the incremental builds that are triggered from application commits, which can help to further identify actual build dependencies. 
\textsc{EChecker} generates only a small number of false positives if \textsc{GNU Make} cannot rebuild a project due to improper build commands. These false positives are relatively rare because they come from improperly compiled commands (\ie command ``ln -s'' \cf \Cref{Effectiveness}) rather than from the design flaws in \textsc{EChecker}.

\sloppy{We evaluate the effectiveness and efficiency of \textsc{EChecker} in 12 representative projects with 240 commits (20 commits per project) and compared it with the latest publicly available state-of-the-art method (Buildfs~\cite{Sotiropoulos20}). We simulate continuous code updates (20 commits) in real-world development by selecting consecutive commits. 
The evaluation results show that \textsc{EChecker} is more effective and efficient at detecting build dependency errors than Buildfs. Specifically, \textsc{EChecker} detects 1,635 MDs and 663 RDs with only 32 false positives. \textsc{EChecker} improves detection efficiency over Buildfs by 85.14 times (with the median at 16.30 times), while significantly reducing the time it takes to detect build dependency errors. \textsc{EChecker} shows its great potential to detect build dependency errors efficiently. As the frequency of commits increases, developers can continue to reap the benefits.}

The main contributions of this paper are as follows.
\begin{itemize}
    \item We propose a novel approach called \textsc{EChecker} to detect build dependency errors. Unlike existing methods, we improve the detection of build dependency errors using incremental builds and avoid time-consuming clean builds when possible.

    \item We present the idea of inferring actual build dependencies based on code changes in commits, which is useful for the detection of build dependency errors. 
    
    \item We evaluate \textsc{EChecker} and demonstrate its significant improvement in the efficiency of detecting dependency errors with satisfactory effectiveness. \textsc{EChecker} can be expected to facilitate software development by efficiently and continuously detecting build dependency errors.
\end{itemize}

\section{Background and Motivation}\label{Background}
This section introduces the basics of build dependency errors and further motivates the issue.

\paragraph{Build dependencies and dependency errors} 
This build process involves analyzing the build scripts that contain the targets that need to be built or not~\cite{GNUmake20}. Dependencies are required for the build of the target and the build commands to generate the target.
In \Cref{FIG:errors_example}, the build target ``BINS'' requires C source files of the format (``src/*.c'') at build time; thus, the build target ``BINS'' is said to depend on dependency ``src/*.c''. Targets and dependencies form a directed acyclic graph, known as a dependency graph~\cite{Actual22}. In this graph, targets are nodes and target dependencies are edges (\ie ``BINS'' is a node, ``src/*.c'' is an edge).

Typically, each build has two kinds of dependencies: declared build dependencies and actual build dependencies~\cite{Actual22}. The actual build dependency graph is derived from the declared build dependencies. The Makefile forms declared build dependencies through targets and prerequisites. Actual build dependencies are needed during the build to produce the correct build of targets---thus, they are a dynamic property; for example, if BINS and all its dependencies are up-to-date, then ``BINS'' will be built correctly. If a dependency is the actual build dependency of the target (``BINS''), the dependency (``src/*.c'') must exist, be constructed, and be updated when the target is built, regardless of whether it is declared as an actual build dependency of the target~\cite{Actual22}. \textsc{GNU Make} will decide to build a minimal set or a re-build target based on declared build dependencies and modification time. An error will be generated if the actual build dependencies do not match the declared build dependencies.

\paragraph{Motivation}
Buildfs~\cite{Sotiropoulos20} is considered the state-of-the-art approach.
Buildfs improves detection efficiency and supports the detection of dependency errors in C and Java projects. Buildfs cuts the build task into a series of build targets and obtains the actual build dependencies for each target by performing a clean build on each target.  
While Buildfs has shown to be highly effective, it has issues in the context of conditional compilation. Conditional compilation refers to the execution of subsequent builds based on the pre-order building or conditions (\eg environment variables and build parameters).
Buildfs ignores the existence of conditional compilation, which can omit or redundantly build artifacts, negatively affecting build efficiency and effectiveness. Since Buildfs builds each target individually, the build system is unable to correctly determine the build conditions. This causes false positives and false negatives to occur during the build process. Furthermore, Buildfs is unable to detect RDs.

Using state-of-the-art approaches like Buildfs, practitioners cannot obtain timely feedback on dependency errors with frequent updates. Typically, they comprehensively test each commit to ensure that it is free from hidden bugs. Thus, ideally, test results would be available immediately after each test. However, in reality, the testing phase is a time-consuming process (\eg it takes hours to detect build dependency errors) and therefore feedback can significantly lag. As a result, developers need to wait longer to see the results of these tests, slowing down their development, and fixing issues potentially only after having started a new task.
To address this challenge, we propose a new approach capable of quickly detecting build dependency errors with improved effectiveness. It can be implemented locally or integrated into Continuous Integration (CI) to help practitioners quickly perform dependency error detection during development.

\vspace{-1.0ex}
\section{Approach}\label{3}
We propose an approach called \textsc{EChecker} to detect build dependency errors. The core idea of \textsc{EChecker} is to efficiently detect build dependency errors by inferring the actual build dependencies. Our key observation is that actual build dependencies of the target change only when the source pre-processor directives and the build commands in the Makefile are changed. Specifically, \textsc{EChecker} infers the actual build dependencies based on changes to pre-processor directives or build commands in the commit. This allows us to obtain new actual build dependencies without performing a clean build. In addition, \textsc{EChecker} monitors incremental builds, which in turn improves the effectiveness of detection by allowing our approach to successfully handle hidden dependencies. \textsc{EChecker} transforms the traditional clean build-based dependency error detection to incremental build-based, which improves detection efficiency while ensuring effectiveness.

\subsection{Overview}
\Cref{fig: Approach Overview} illustrates how \textsc{EChecker} works. \textsc{EChecker} takes the project, commit, and the historical actual build dependency graph of clean builds as inputs. 
Assuming that the historical actual build dependency graph of clean builds exists, \textsc{EChecker} detects build dependency errors in three stages. First, \textsc{EChecker} applies changes of a commit to trigger an incremental build and monitors the execution of the incremental builds to obtain the actual build dependency graph of incremental builds (step \textcircled{1}). Second, \textsc{EChecker} analyzes the specifics of the commit. \textsc{EChecker} focuses on whether the commit changes the pre-processor directives of source files and the Makefile (steps \textcircled{2}--\textcircled{4}). Then, \textsc{EChecker} infers the actual build dependency graph by leveraging the results of step \textcircled{2}--step \textcircled{4} (step \textcircled{5}). Finally, \textsc{EChecker} detects build dependency errors based on the updated actual build dependency graph (step \textcircled{6}). If the historical actual build dependency graph of clean builds does not exist, \textsc{EChecker} performs a clean build to detect dependency errors and saves the actual build dependency graph as input for the next detection. Note that, for the detection of multiple commits merely one clean build is required for \textsc{EChecker}. Then it will continuously update the actual build dependency graph by inferring actual build dependencies.

\subsection{Inferring Actual Build Dependencies}\label{sec:inferring actual build dependencies}
This subsection describes in detail how \textsc{EChecker} infers the actual build dependency graph in three stages: monitoring incremental builds execution, pre-processor directives, and Makefile change analysis, and inferring actual build dependencies.
\begin{figure}[!b]
  \vspace{-2.0ex}
  \centering
  \includegraphics[scale=0.5,trim=0 435 480 0,clip]{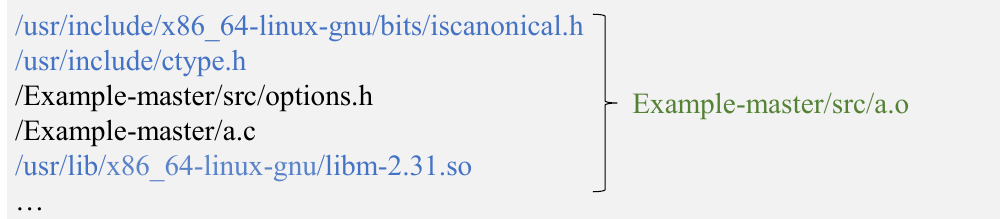}
  \caption{Tracing Build of the Example Projects}
  \label{fig: Build Trace from Example Projects}
\end{figure}
\subsubsection{Monitoring Incremental Builds Execution} \textsc{EChecker} monitors the execution of the incremental builds to obtain the actual build dependency graph of incremental builds (step \textcircled{1}).

\begin{figure*}[!t]
  \centering
  \includegraphics[scale=0.523,trim=0 235 0 0,clip]{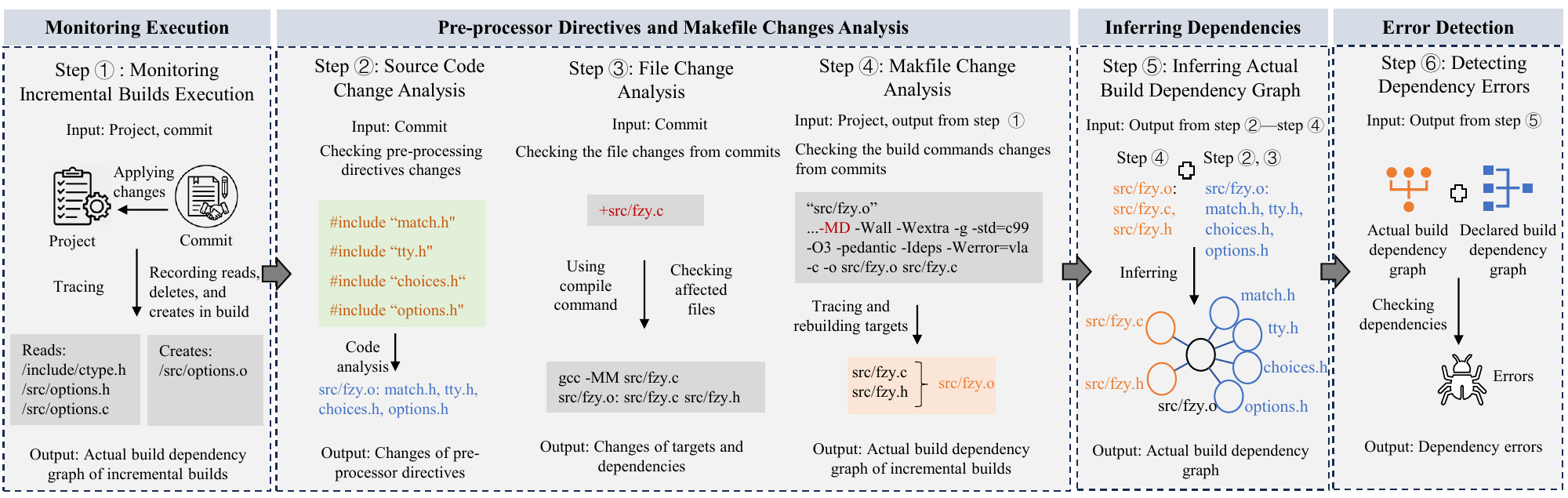}
  \caption{Overview of \textsc{EChecker}}
  \label{fig: Approach Overview}
  \vspace{-3.0ex}
\end{figure*}

The actual build execution information has to be retrieved by tracing the build process executed by \textsc{GNU Make}. In a file system-based build, the build is composed of a series of operations on files, \eg \textit{reads}, \textit{deletes}, and \textit{creates}~\cite{systemcall,communication}. The build execution information is implicitly exposed in the system calls. \textsc{GNU Make} creates a sub-process for each build target and executes the corresponding build commands in the sub-process. The execution of the build commands is essentially the process of manipulating the underlying files. Therefore, it is possible to trace the operations performed on the underlying files during the build process to obtain the actual build dependencies for each build target. Based on the type and parameter information of all captured system calls, the files manipulated by each process can be divided into input files (files called during the build process, \ie build process reads ``/src/options.h'' in \Cref{fig: Approach Overview}) and output files (files newly created or written during the build process, \ie build process creates ``/src/options.o'' in \Cref{fig: Approach Overview}). By monitoring the build process, \textsc{EChecker} can obtain actual build dependencies.

\textsc{EChecker} is designed to report potential dependency errors only to the project, not external dependencies, such as standard libraries.
\textsc{EChecker} excludes external dependencies of the project by file paths. As shown in \Cref{fig: Build Trace from Example Projects}, \textsc{EChecker} excludes the files (blue text in the figure) by their file path, not the external dependencies (not in the project file path, \eg ``/Example-master/'' in the figure). For example, most projects rely on the standard library (\eg ``/x86\_64-linux-gnu/bits/iscanonical.h'' in the figure). \textsc{EChecker} may report a large number of missing basic libraries if they are not excluded. In reality, it is difficult for practitioners to include all external libraries needed in a Makefile. Many tools (\eg \textsc{Autotools}~\cite{auto20} and \textsc{CMake}\cite{CMake10}) are designed to help practitioners automatically generate a Makefile without having to consider external libraries.

\subsubsection{Pre-processor Directives and Makefile Change Analysis} Monitoring incremental builds helps to obtain hidden dependencies. However, the actual build dependency graph for incremental builds is not complete enough, because the execution of incremental builds can be affected by dependency errors that we aim to find. For example, the incremental builds system does not perform re-building of the source files when the header files in~\Cref{FIG:errors_example} are changed. The idea of \textsc{EChecker} is that the incomplete actual build dependency graph obtained by monitoring can be complemented by analyzing the pre-processor directives and the Makefile, which is divided into two phases: pre-processor directives change analysis and build commands change analysis (steps \textcircled{2}--\textcircled{4}).

\paragraph{Pre-processor directives change analysis} The changes from commits include changes to pre-processor directives and source files. 

\begin{figure}[!b]
  \vspace{-2.0ex}
  \centering
  \includegraphics[scale=0.5,trim=0 405 480 0,clip]{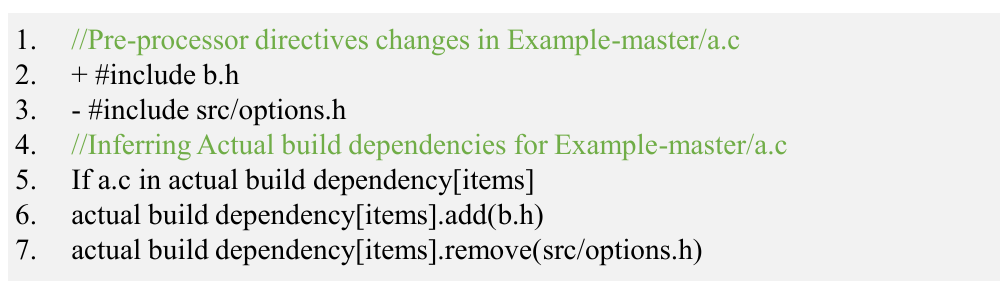}
  \caption{Inferring Actual Build Dependencies for Changes to Pre-processor Directives}
  \label{CODE:inferring actual build dependency for Changes to Pre-processor Directives}
\end{figure}

For pre-processor directives, \textsc{EChecker} identifies changes made to the pre-processor directives by a commit (\eg additions and deletions). All nodes that depend on a file are then updated in the actual build dependency graph, and actual build dependencies are added and removed based on the changes. 
At build time, \textsc{GNU Make} recognizes the files referenced by the source file (files beginning with the ``\#include'') and includes those files in the build of the source file. 
Files starting with ``\#include'' are considered to be actual build dependencies. \Cref{CODE:inferring actual build dependency for Changes to Pre-processor Directives} illustrates how \textsc{EChecker} updates the actual build dependencies when the pre-processor directives are changed. The pre-processor directives of ``/Example-master/a.c'' are removed and added respectively (lines 2 and 3). \textsc{EChecker} updates each node containing ``/Example-master/a.c'' in the actual build dependency graph (lines 5--7).

\begin{figure}[!b]
  \vspace{-2.0ex}
  \centering
  \includegraphics[scale=0.5,trim=0 340 480 0,clip]{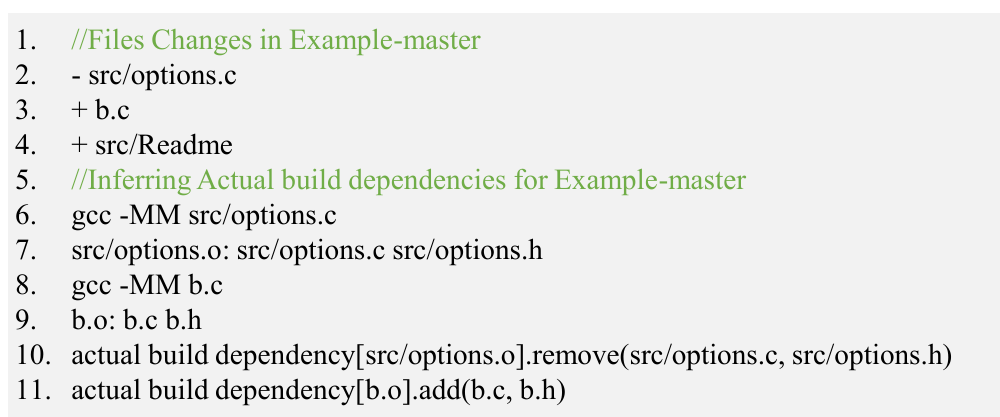}
  \caption{Inferring Actual Build Dependencies for Changes to Files}
  \label{CODE:inferring actual build dependencies for Changes to Files}
\end{figure}

\begin{figure}[!b]
  \vspace{-4.0ex}
  \centering
  \includegraphics[scale=0.5,trim=0 385 480 0,clip]{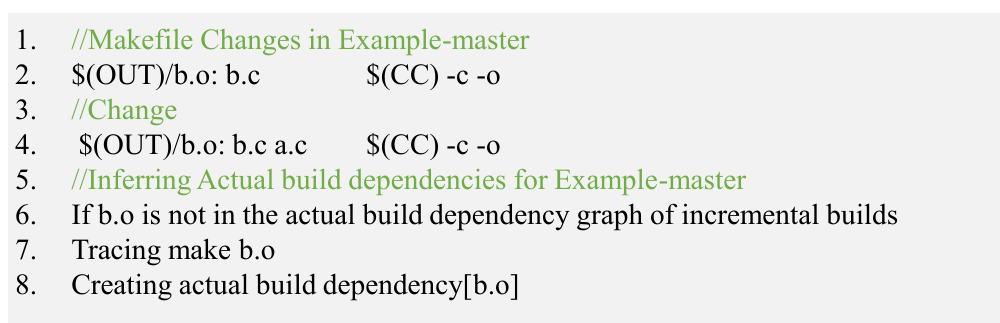}
  \caption{Inferring Actual Build Dependencies for Changes to Makefile}
  \label{CODE:inferring actual build dependencies for Changes to Makefile}
\end{figure}

For the files that are added or deleted, \textsc{EChecker} checks their file types by the suffix to determine subsequent actions. For source code types (\eg ``.c'', ``.cpp'', ``.cc'', ``.h'', ``.hpp''), \textsc{EChecker} uses the compiler commands (\eg gcc -MM)~\cite{GCC} to obtain the files on which the added or deleted files depends and then add or remove these files (nodes) from the actual build dependencies.
The files that have a non-source file type are updated during the incremental build monitoring step and do not need to be processed again here. For renamed files, \textsc{EChecker} uses the new filename to replace the original actual build dependencies.
\Cref{CODE:inferring actual build dependencies for Changes to Files} is an illustration of \textsc{EChecker} infers the actual build dependencies when the files are changed. The project ``Example-master'' removed a file and added two files (lines 2--4). \textsc{EChecker} infers the ``src/options.o'' and ``b.o'' in the actual build dependency (lines 6--11). The file ``src/Readme'' is not a source code file, \textsc{EChecker} will ignore the addition of the file ``src/Readme'' (line 4).

\paragraph{Build command change analysis} \textsc{EChecker} instructs \textsc{GNU Make} to output the build commands for all build targets (using command ``make -n -B --debug=basic'') after applying the commit and to use all build commands as input for the next detection. \textsc{EChecker} uses a lightweight comparison to analyze whether the build commands of targets have changed. \textsc{EChecker} identifies whether the target whose build command was changed was rebuilt during the previous incremental build. If not, the build target is rebuilt separately to obtain its actual build dependencies.
\Cref{CODE:inferring actual build dependencies for Changes to Makefile} illustrates how \textsc{EChecker} infers the actual build dependencies when the Makefile is changed.
The prerequisite ``a.c'' (lines 1--5) is added to the target ``\$(OUT)/b.o''. \textsc{EChecker} checks if ``b.o'' is in the actual build dependency graph of incremental builds. If not, \textsc{EChecker} rebuilds ``b.o'' and traces the build process to obtain its actual build dependencies (lines 6--8).

\subsubsection{Inferring Actual Build Dependency Graph} \textsc{EChecker} infers a new actual build dependency graph by leveraging the results of the monitoring incremental builds execution stage and the static code analysis stage, and the historical actual build dependency graph of clean builds (step \textcircled{5}).

\textsc{EChecker} traverses all the targets from the previous two stages. For all targets in both graphs, \textsc{EChecker} checks in the historical actual build dependency graph of clean builds for targets, and updates them if they exist. If targets do not exist in the history actual build dependency graph of clean builds, \textsc{EChecker} adds them to the history actual build dependency graph. After a traversal, \textsc{EChecker} obtains the new actual build dependency graph.

\subsection{Detecting Build Dependency Errors}
\textsc{EChecker} detects dependency errors by leveraging the actual build dependency graph and the declared build dependency graph. In this subsection, we illustrate how \textsc{EChecker} obtains a declared build dependency graph and detects dependency errors (step \textcircled{6}).

\begin{figure}[!t]
  \centering
  \includegraphics[scale=0.5,trim=0 450 480 0,clip]{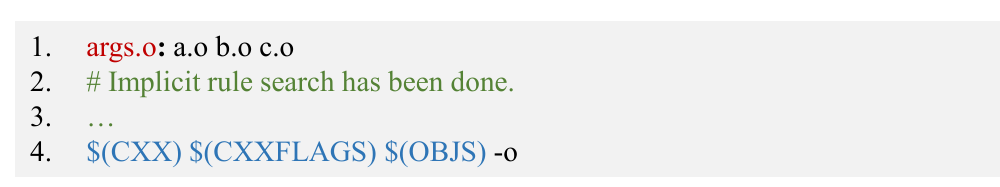}
  \caption{Date Structure of the Internal Database}
  \label{fig: Data Structure}
  \vspace{-4.0ex}
\end{figure}

\paragraph{Obtaining declared build dependency graph}
\textsc{EChecker} instructs \textsc{GNU Make} to output its internal database (command ``make -p'')~\cite{GNUmake20} to obtain the declared build dependencies. \textsc{GNU Make} parses the Makefile during the build and stores the resulting declared build dependencies in its internal database. \textsc{EChecker} obtains the declared build dependency graph by parsing the internal database of \textsc{GNU Make} (line 1 in \Cref{fig: Data Structure}). \textsc{GNU Make} will output \textit{Phony Targets}, which are targets that do not represent actual targets (\eg phony target ``all''), but are simply a mechanism for organizing and managing target dependencies~\cite{GNUmake20}. Typically, the name of a phony target is user-defined and declared in the ``.PHONY'' variable in Makefile. Phony targets in the declared build dependency graph will cause false positives (reporting a dependency error as a dependency error when it is not), hence \textsc{EChecker} identifies the phony targets and replaces them with the value of the ``.PHONY'' variable (the real targets) to avoid reporting false positives.

\paragraph{Detecting build dependency errors}
\textsc{EChecker} detects build dependency errors by comparing the actual build dependency graph and the declared build dependency graph.
For the same targets, MDs are in the actual build dependency graph, but not in the declared build dependency graph. RDs are the dependencies in the declared build dependency graph that do not appear in the actual build dependency graph.

\section{Illustrative Example}

In this section, we use a concrete example to illustrate how \textsc{EChecker} infers the actual build dependencies and thus detects dependency errors in the \textit{fzy} project,
a fast, simple fuzzy text selector for the terminal. Commit 28195b3 from the \textit{fzy} project has two MDs (``src/match.h'' and ``src/tty.h'') in the target (``src/fzy.o'').

\begin{figure}[!b]
  \vspace{-3.0ex}
  \centering
  \includegraphics[scale=0.5,trim=0 385 480 0,clip]{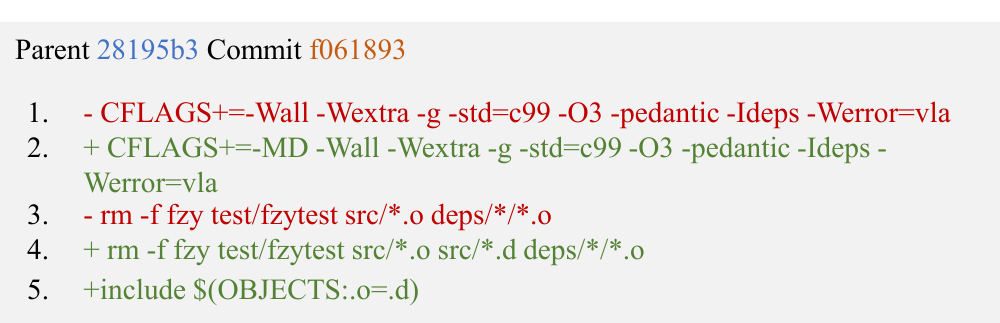}
  \caption{Commit from \textit{Fzy} Project}
  \label{FIG:code change}
\end{figure}
\paragraph{Current state} Assume that \textsc{EChecker} is currently obtaining the actual build dependency graph from the parent commit (commit ID: 28195b3) since the practitioner has been using \textsc{EChecker} to check the past commits. If not, \textsc{EChecker} would obtain it through a clean build. \textsc{EChecker} first obtains the historical actual build dependencies of all targets through the historical actual build dependency graph.

\begin{figure}[!t]
  \centering
  \includegraphics[scale=0.5,trim=0 500 480 0,clip]{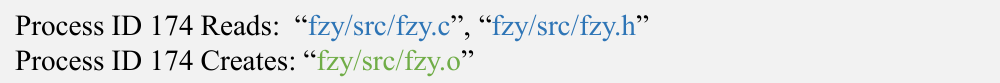}
  \caption{Tracing Build from \textit{Fzy} Project}
  \label{fig:Trace Build to Create Actual Graph}
  \vspace{-3.0ex}
\end{figure}

\paragraph{Monitoring incremental builds execution}
 \textsc{EChecker} applies the changes from the commit (commit ID: f061893, as shown in \Cref{FIG:code change}), executes an incremental build, and traces it (step \textcircled{1} in ~\Cref{fig: Approach Overview}). \textsc{EChecker} identifies each file output by \textsc{GNU Make} and each file input based on file system calls, obtaining the actual build dependency graph of incremental builds between the files (build targets). As~\Cref{fig:Trace Build to Create Actual Graph} shows, following the process (ID 174), the process (ID 174) reads the ``src/fzy.c'' and ``src/fzy.h'', and generates ``src/fzy.o''. Therefore, \textsc{EChecker} creates a node named ``src/fzy.o'', which has two edges (``src/fzy.c'' and ``src/fzy.h''). 

\paragraph{Pre-processor directives and Makefile change analysis}
\textsc{EChecker} determines whether to update the actual build dependencies of the targets by analyzing the pre-processor directives in the source code and the build commands in the Makefile ((steps \textcircled{2}--\textcircled{4} in ~\Cref{fig: Approach Overview})). 
The idea of \textsc{EChecker} is to build on top of the actual build dependency graph of incremental builds and then complete the actual build dependency graph with code analysis. 
\textsc{EChecker} first identifies the pre-processor directives changes of commits in source files. 
If new pre-processor directives occur, it infers the new pre-processor directives as new actual build dependencies. Otherwise, \textsc{EChecker} considers the actual build dependencies of the target to be consistent with the historical actual build dependencies. 
In addition, \textsc{EChecker} analyzes the files for changes (\eg deletions or additions). No file changes are involved in this example. 
More details are presented in \Cref{sec:inferring actual build dependencies}.

\begin{figure}[!t]
  \centering
  \includegraphics[scale=0.5,trim=0 370 480 10,clip]{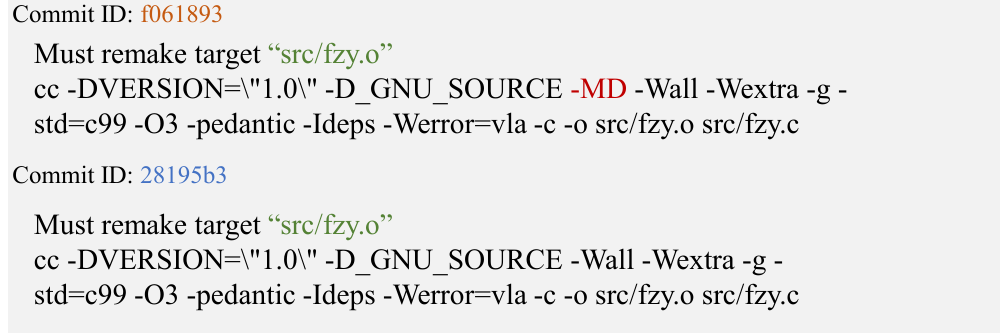}
  \caption{Changes of Build Command from \textit{Fzy} Project with New Command Highlighted in Red.}
  \label{fig: Build Command Change in Makefile from Fzy Project. New Command is in Red.}
  \vspace{-3.0ex}
\end{figure}

Second, \textsc{EChecker} identifies whether the build commands in the Makefile have changed. 
It instructs \textsc{GNU Make} to rebuild all targets whose build commands have changed and to monitor the execution of the build for the actual build dependencies. \Cref{fig: Build Command Change in Makefile from Fzy Project. New Command is in Red.} shows that the ``src/fzy.o'' build commands have changed since it was applied to the commit. \textsc{EChecker} will re-build ``src/fzy.o'' and obtain the actual build dependencies for the new ``src/fzy.o''.

\begin{figure}[!t]
  \centering
  \includegraphics[scale=0.5,trim=0 350 480 0,clip]{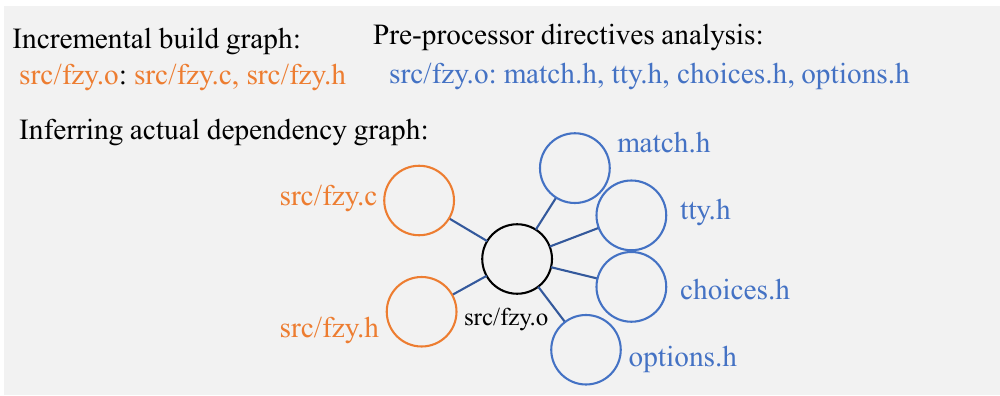}
  \caption{Inferring Actual Build Dependency Graph}
  \label{fig: inferring actual build dependency Graph}
  \vspace{-2.0ex}
\end{figure}

\paragraph{Inferring actual build dependency graph}
As shown in ~\Cref{fig: inferring actual build dependency Graph}, the actual build dependency of ``src/fzy.o'' in the actual build dependency graph of incremental builds has two edges (``src/fzy.c'' and ``src/fzy.h''). 
\textsc{EChecker} infers the actual build dependencies by analyzing the pre-processor directives of the source files affected by the current commit (step \textcircled{5} in \Cref{fig: Approach Overview}). 
By analyzing the pre-processor directives of ``src/fzy.c'' and ``src/fzy.h'', \textsc{EChecker} considers the actual build dependencies the same as historical actual build dependencies. The historical actual build dependencies of ``src/fzy.o'' also contain ``match.h'', ``tty.h'', ``choices.h'', and ``choices.h''. Therefore, \textsc{EChecker} infers that ``src/fzy.o'' actually depends on ``src/fzy.c'', ``src/fzy.h'', ``match.h'', ``tty.h'', ``choices.h'', and ``options.h''.

\paragraph{Detecting build dependency errors}
\textsc{EChecker} detects dependency errors by leveraging the actual build dependency graph and the declared build dependency graph (step \textcircled{6} in \Cref{fig: Approach Overview}).
It instructs \textsc{GNU Make} to output declared build dependencies and check that the actual build dependency of each target matches the declared build dependency; any mismatch will be reported as a dependency error. For example, ``src/fzy.o'' has six actual build dependencies (\cf \Cref{fig: inferring actual build dependency Graph}), but the declared build dependencies of ``src/fzy.o'' only have ``src/fzy.c'', ``src/fzy.h'', ``choices.h'', and ``options.h''. Thus, ``match.h'' and ``tty.h'' are MDs of ``src/fzy.o''.

\section{Evaluation}\label{5}
In our evaluation, we seek to explore the effectiveness and efficiency of \textsc{EChecker} in real-world software projects. Buildfs~\cite{Sotiropoulos20} is the state-of-the-art available method for detecting build dependency errors. 
Buildfs detects MDs and ordering violations (not specify ordering constraints between two dependent tasks) for C projects, but not RDs. For an apples-to-apples comparison, we configure Buildfs to detect only MDs and not execute ordering violations. We are not concerned with ordering violations as same with other methods~\cite{Fan20,Wu22}.
We propose the following questions (Q).
\begin{itemize}
    \item \textbf{Q.1}: How effective is \textsc{EChecker} compared to Buildfs in detecting build dependency errors?
    \item \textbf{Q.2}: How efficient is \textsc{EChecker} compared to Buildfs in detecting build dependency errors?
\end{itemize}

To answer these questions, we used \textsc{EChecker} and Buildfs to detect build dependency errors in 12 projects with 240 commits. We used Buildfs' results of detecting build dependency errors and ground truth as the baselines. For effectiveness, we evaluated whether \textsc{EChecker} can detect as many or even more build dependency errors as Buildfs (Q.1.1). We calculated precision, recall, and F-1 score for each approach based on ground truth (Q.1.2). In addition, we reported our detection results to the maintainer of projects and evaluated whether the results of our detection were accepted by the maintainers (Q.1.3).
For efficiency, we compared the time consumption of \textsc{EChecker} and Buildfs for each stage of dependency error detection. Note that for a fair comparison, the time consumption of \textsc{EChecker} includes the time to obtain a historical actual build dependency graph of clean build. In addition, we established two baselines:(1) the first baseline is all 200 commits, and (2) the second baseline is 68 commits with changes to build scripts and pre-processor directives. Since a commit does not generate a new dependency error if practitioners do not modify build scripts or pre-processor directives, it is not necessary to detect build dependency errors in the commit. We manually checked each commit on the first baseline (240 commits) to obtain the second baseline (68 commits). In the second baseline, we did not include \textit{OpenCV} and \textit{redis} because Buildfs cannot detect build dependency errors in them.

\subsection{Experimental Settings}
\paragraph{Subjects}
We selected 12 projects based on the following three characteristics: they use \textsc{GNU Make} or \textsc{CMake} as the build system, they are highly mature and popular projects, and have a significant project size (large (\textgreater1000) / medium (100--1000) / small (\textless100) number of files)~\cite{Fan20,Wu22}. From these projects, we chose 20 consecutive commits, meaning that we overall obtained 240 commits. We randomly selected a commit as the first of 20 commits and selected the next 19 commits in chronological order. The first commit was replaced if there were fewer than 19 commits following the first commit.
Important information about subjects is shown in \Cref{table:Important Information about the Subjects}.

\begin{table}[!t]
	\centering
	\caption{Important Information about Subjects}
    \fontsize{7.5pt}{10pt}\selectfont
	\begin{tabular}{llrrrl}
		\toprule
		System & Name & Stars & Files & Commits & Commit ID Range\\
		\midrule
		\multirow{6}{*}{\textsc{GNU Make}} & fzy & 6.1k & 16  & 453 & \textcolor{teal}{eb4cbd4--f061893}\\
        ~& chibicc &  7.4k & 77 & 316 & \textcolor{teal}{c0f0614--90d1f7f}\\
		~ & ck & 2.2k & 501  & 1,654 & \textcolor{teal}{dac27da--7eff0db}\\
        ~ & clib & 4.5k & 250 & 434 & \textcolor{teal}{c7c5ff6--4390598}\\ 
        ~ & redis & 59.5k & 1,465 & 11,679 & \textcolor{teal}{d2d6bc1--395d801}\\ 
		~ & python & 50k & 4443  & 115,763 & \textcolor{teal}{3c87a66--b4612f5}\\
  \midrule
        \multirow{6}{*}{\textsc{CMake}} & libco & 7.4k & 28  & 103 & \textcolor{teal}{580a446--dc6aafc}\\
        ~ & cJSON & 8.3k & 228  & 1,073 & \textcolor{teal}{d44b594--c680fae}\\
        ~ & fastText & 24.1k & 526  & 381 & \textcolor{teal}{231b871--3697152}\\
        ~ & gravity & 4.2k & 1,247 & 764 & \textcolor{teal}{ee4a0b0--0feea4b} \\
        ~ & cppcheck & 4.5k & 1,248  & 26,579 & \textcolor{teal}{e21dca2--dc0b352}\\
       ~ & OpenCV & 72k & 10,644  & 33,466 & \textcolor{teal}{b3d3acf--c96f48e}\\ 
		\bottomrule
	\end{tabular}
	\label{table:Important Information about the Subjects}
 \vspace{-5.0ex}
\end{table}

\paragraph{Approach application methodology}
All subjects are built using their default configuration. After executing the pre-processing of the project (\eg \texttt{./autogen}, \texttt{./configure}, and \texttt{./cmake}), Buildfs and \textsc{EChecker} check each project separately. To ensure that Buildfs and \textsc{EChecker} do not interfere with each other, we executed them in separate environments (in separate docker containers). 
For \textsc{EChecker}, we first performed a clean build on the first commit (\eg b3d3acf in \textit{OpenCV}) to obtain the actual build dependency graph and included this time in the \textsc{EChecker}'s time consumption. We then performed incremental builds for detecting dependency errors by applying changes from the new commits in \textsc{EChecker} detection. For Buildfs, we performed a clean build to detect dependency errors in each clean environment. In addition, we reported the dependency errors detected in nine projects to the maintainers, of which we obtained positive feedback from the maintainers of two projects (\textit{cppcheck}, \textit{cJSON}) who confirmed our reports.

\paragraph{Ground truth} We calculated the ground truth based on build procedures and user declarations for each subject with the default configuration. First, we parsed the build script for declared build dependencies and traced the build procedures to obtain the actual build dependencies. The declared build dependencies that do not match the actual build dependencies are identified as MDs. Moreover, for further validation of MDs, we modified the timestamps of the MDs one by one and triggered the incremental builds. There are true MDs which are expected files (the build targets of MDs) that are not rebuilt. The reason for this is that modifications to MDs do not cause a rebuild of the target file to occur~\cite{Licker19}. Based on these, we obtain the ground truth for MDs. For RDs, we systematically and automatically removed non-actual build dependencies from the declared build dependencies for each build target and then built the target individually. The dependencies are true RDs if the target can still be successfully built after the dependency removal. Since RDs are not required for building the targets, they are just the dependencies of the practitioner's redundant declarations in Makefile~\cite{Licker19}. Therefore, removing them will not affect the target building process. We consider a true positive (TP) if \textsc{EChecker} agrees with the ground truth, a false positive (FP) if it does not, and a false negative (FN) if the error is not reported by \textsc{EChecker}.
Nevertheless, no previous study in this area ever provide a ground truth~\cite{Bezemer17,Licker19,Sotiropoulos20,Fan20,Wu22}.

\paragraph{Experiment environment} Experiments were conducted on a Linux server running Ubuntu 22.04 with an Intel(R) Xeon(R) Platinum 8163 CPU@2.50GHz, 96 cores, and 376GB of physical memory. 
To eliminate experimental bias due to variations in device performance, we averaged the time consumed on detecting dependency errors of the two approaches by performing each detection twice.

\subsection{Evaluation of Effectiveness (Q.1)}\label{Effectiveness}
The results of detected errors are shown in Table~\ref{tab:Detecting dependency errors in project}. The six columns show the project name, the number of MDs and RDs detected by the two approaches (Buildfs and \textsc{EChecker}), the number of confirmed (overlapped) MDs, and the new dependency errors found by \textsc{EChecker}. In the ''Confirmation'' column, we listed both the same (''MDs'') and different (''New'') build dependency errors between Buildfs and \textsc{EChecker}. In the column, we excluded false positives of \textsc{EChecker}. For example, in the \textit{cJSON} project, we found 17 MDs, 16 of which were false positives of \textsc{EChecker}. Therefore, after excluding the false positives, \textsc{EChecker} found one error, the same as Buildfs and 0 error reports were newly found build dependency errors in the \textit{cJSON} project. 
\begin{table}[!t]
    \centering
     \caption{Detecting Dependency Errors in Projects. Buildfs Does not Support Detecting RDs.}
    \fontsize{7.5pt}{10pt}\selectfont
    \begin{threeparttable}
    \vspace*{-1.0ex}
   \begin{tabular}{lrrrrr}
    \toprule
        \multirow{2}{*}{Subject} &\multicolumn{1}{c}{Buildfs}  & \multicolumn{2}{c}{\textsc{EChecker}}  & \multicolumn{2}{c}{Confirmation}  \\ 
        ~ & \multicolumn{1}{c}{(FPs) MDs} & (FPs) MDs & (FPs) RDs & MDs & New \\
    \midrule    
        chibicc & 0 & 0 & 0 & 0 & 0 \\ 
        cJSON & 1 & (16) 17 & (16) 16 & 1 & 0 \\ 
        ck & (169) 385 & 216 & 0 & 216 & 0 \\ 
        clib & 94 & 94 & 0 & 0 & 0 \\ 
        cppcheck & 18 & 18 & 0 & 18 & 0 \\ 
        fastText & 35 & 35 & 6 & 35 & 0 \\ 
        fzy & 17 & 17 & 1 & 17 & 0 \\ 
        gravity & 287 & 287 & 0 & 287 & 0 \\ 
        libco & 36 & 36 & 0 & 36 & 0 \\ 
        redis & n$\backslash$a & 0 & 0 & n$\backslash$a & n$\backslash$a \\ 
        python & (18) 727 & 915 & 640 & 709 & 206 \\ 
        OpenCV & n$\backslash$a & 0 & 0 & n$\backslash$a & n$\backslash$a \\
        \midrule
        Sum & (187) 1,600 & (16) 1,635 &(16) 663 & 1,413 & 206 \\ 
        \bottomrule
    \end{tabular}
      \begin{tablenotes}
        \fontsize{7pt}{7.5pt}\selectfont
        \item FPs: False Positives
     \end{tablenotes}
     \end{threeparttable}
    \vspace{-3.0ex}    
    \label{tab:Detecting dependency errors in project}
\end{table}

\paragraph{Results and true positives} \textsc{EChecker} detects 1,635 MDs and 663 RDs in 12 projects with 240 commits. Buildfs detects 1,600 MDs. The overlap in terms of actual errors between \textsc{EChecker} and Buildfs is 1,413 MDs, with 206 new MDs found. \textsc{EChecker} confirms the MDs detection results from Buildfs on 180 commits in 7 projects. \textsc{EChecker} can detect more dependency errors than Buildfs in 4 projects with two projects (\textit{OpenCV} and \textit{redis}) in which Buildfs cannot detect errors. Based on the ground truth of the subjects, \textsc{EChecker} has 32 false positives and 0 false negatives. On the contrary, Buildfs has 187 false positives and 206 false negatives. The precision of \textsc{EChecker} is 99.1\% and that of Buildfs is 88.3\%. The recall of \textsc{EChecker} is 100\% and that of Buildfs is 75.7\%. The F-1 score for \textsc{EChecker} is 0.995 and for Buildfs it is 0.815. \textsc{EChecker} improves on Buildfs by 10.8\%, 24.3\%, and 0.18 in terms of precision, recall, and F-1 score, respectively.

\begin{figure}[!t]
  \centering
  \includegraphics[scale=0.5,trim=0 380 480 0,clip]{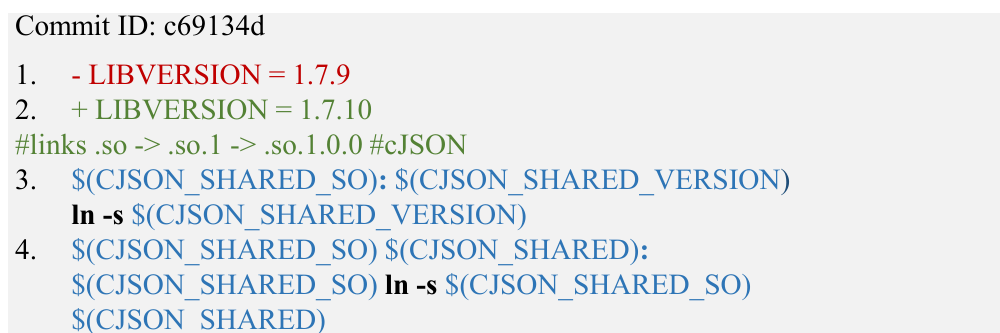}
  \caption{False Positives of \textit{cJSON} Project}
  \label{FIG:FPs}
  \vspace{-4.0ex}
\end{figure}

\paragraph{False positives} \textsc{EChecker} reports 32 false positives (16 MDs and 16 RDs) in \textit{cJSON} (commit ID: c69134d, 09ebae8, 93688cb, 687b1a2).
The key reason for the false positives is the improper use of the soft link command in the build command of the target (as shown in \Cref{FIG:FPs}). A deeper analysis of the false positives reveals that the issue occurs when the target ``libcjson.so.1'' is being built. The build command is ``ln -s'', which builds ``libcjson.so.1'' via a soft link to ``libcjson'' (lines 3 and 4). The build command uses \textsc{GNU Make} to build ``libcjson.so.1'' by soft linking ``libcjson.so.1.7.09''. However, when ``libcjson.so.1'' is present, the build command ``ln -s'' causes \textsc{GNU Make} not to retry creating the soft link. This means that when the ``LIBVERSION'' is changed (\eg lines 1 and 2), the incremental builds do not complete properly. Thus, \textsc{EChecker} creates an incorrect actual build dependency graph. However, since the declared build dependencies in the Makefile have been changed in the new version, the wrong actual build dependency graph leads to false positives for MDs on the old linked source file (``libcjson.so.1.7.09''). As a result, \textsc{EChecker} produces false positives for the new linked source file (``libcjson.so.1.7.10'') and redundant dependencies on new versions of linked sources (``libcjson.so.1.7.10''). A fix is to change the link command to ``ln -sf'' so that the file will be regenerated even if the soft-link file already exists. 
In other commits in \textit{cJSON} project, no further false positives are generated as the version ``LIBVERSION'' has not changed.
\paragraph{False negatives} \textsc{EChecker} does not generate false negatives in the configuration being tested.
\textsc{EChecker} can only detect dependency errors in the current build configuration and will miss dependency errors in other build configurations. Practitioners usually build different software variants through software configuration. The solution is to detect software variants sequentially. Buildfs' false negatives result from the inability to properly perform conditional compilation and correctly identify the build target (\eg build target ``Modules/grpmodule.o'' in \textit{python}). Buildfs obtains build targets by correctly simulating build execution, but for many projects correctly simulating build execution before a clean build is not feasible (\eg \textit{python}). In addition, Buildfs is unable to detect \textit{redis} and \textit{OpenCV}. It fails to obtain the build targets for these two projects since it cannot simulate the execution of these two projects due to the lack of pre-order artifacts. Thus, Buildfs is unable to detect dependency errors in the two projects.

\sloppy{\paragraph{False positives in Buildfs} We observed that for four projects, Buildfs can detect different errors from the errors reported by \textsc{EChecker}. We manually analyzed these reports. Buildfs can detect more errors in the \textit{ck} project. We find that Buildfs reports dependency errors from phony targets (\ie phony target ``all''), which are false positives. Phony targets are characterized by the fact that they do not create a file, but only execute certain commands~\cite{PhonyTarget}. The phony target ``all'' specifies which targets (files) should be built.} 

\vspace{1ex}
\begin{mdframed}[backgroundcolor=light-gray]
\small
\emph{\textbf{Answer to Q.1:} According to the ground truth, \textsc{EChecker} detects more dependency errors than Buildfs (Q.1.1). \textsc{EChecker} improved its F-1 score compared to Buildfs by 0.18 (Q.1.2). The maintainers of the two projects accepted our error reports (Q.1.3), which demonstrates the effectiveness of \textsc{EChecker}.}
\end{mdframed}

\subsection{Evaluation of Efficiency (Q.2)}
We have evaluated the efficiency of \textsc{EChecker} and Buildfs by measuring the time required to detect errors in the aforementioned 12 projects with 240 commits (as shown in~\Cref{tab:time consumption}). For an apples-to-apples comparison, we allowed Buildfs to detect only build dependency errors. The workflow of \textsc{EChecker} and Buildfs consists mainly of tracing builds and checking dependencies. Therefore, the efficiency in the evaluation consists of two components, namely build time and dependency error analysis time. Table \ref{tab:time consumption} shows the build time monitored by \textsc{EChecker} and Buildfs. The table includes a comparison of \textsc{EChecker} and Buildfs in terms of tracing time, detecting build dependency errors time, and sum time. In~\Cref{tab:time consumption}, each value is derived from the median and sum of the 20 commits detected in each project.

\begin{table*}[!t]
    \centering
     \caption{Time Consumption on Detecting Build Dependency Errors in 240 Commits (Sec.)}
    \fontsize{6.2pt}{10pt}\selectfont
    \begin{threeparttable}
    \begin{tabular}{l|rrrrrr|rrrrrrrrrr|rr}
    \toprule
        \multirow{3}{*}{Projects} &  \multicolumn{6}{c|}{Buildfs}  & \multicolumn{10}{c|}{\textsc{EChecker}} &\multicolumn{2}{c}{\multirow{2}{*}{Sum Difference}} \\ 
        \cline{2-17}
        ~ & \multicolumn{2}{c}{Trace} & \multicolumn{2}{c}{Detection} &\multicolumn{2}{c|}{Sum} & \multicolumn{2}{c}{Trace} &\multicolumn{2}{c}{IBGraph} &\multicolumn{2}{c}{Inferring} &\multicolumn{2}{c}{Detection} & \multicolumn{2}{c|}{Sum} & ~ & ~\\ 
        ~ & \multicolumn{1}{c}{Med} & \multicolumn{1}{c}{All} & \multicolumn{1}{c}{Med} & \multicolumn{1}{c}{All} & \multicolumn{1}{c}{Med} & \multicolumn{1}{c|}{All} & \multicolumn{1}{c}{Med} & \multicolumn{1}{c}{All} & \multicolumn{1}{c}{Med} & \multicolumn{1}{c}{All} & \multicolumn{1}{c}{Med} & \multicolumn{1}{c}{All} & \multicolumn{1}{c}{Med} & \multicolumn{1}{c}{All} & \multicolumn{1}{c}{Med} & \multicolumn{1}{c|}{All} & 
        \multicolumn{1}{c}{Med} & \multicolumn{1}{c}{All} \\ \midrule
        chibicc & 3.46 & 68.51 & 0.02 & 0.43 & 3.48 & 68.94 & 0.68 & 25.7 & 0.03 & 0.71 & 0.07 & 2.18 & 0.01 & 0.12 & 0.79 & 28.71 & 2.69 & 40.23 \\ 
        cJSON & 1.72 & 33.73 & 0.02 & 0.49 & 1.74 & 34.22 & 0.42 & 7.9 & 1.05 & 21.06 & 0.07 & 1.27 & 0.01 & 0.12 & 1.60 & 30.35 & 0.14 & 3.87 \\ 
        ck & 4.68 & 94.26 & 0.10 & 1.91 & 4.78 & 96.17 & 0.24 & 10.37 & 0.03 & 0.54 & 0.03 & 2.23 & 0.01  & 0.20  & 0.40 & 13.34 & 4.38 & 82.83 \\ 
        clib & 29.22 & 576.02 & 0.25 & 4.96 & 29.48 & 580.97 & 0.12 & 165.43 & 0.06 & 5.16 & 3.25 & 64.92 & 0.05 & 1.00  & 3.64 & 236.51 & 25.84 & 344.46 \\ 
        cppcheck & 257.22 & 5,115.17 & 0.12 & 2.34 & 257.34 & 5,117.51 & 0.11 & 623.73 & 0.06 & 1.56 & 0.11 & 4.88 & 0.09 & 2.04 & 0.39 & 632.21 & 256.95 & 4,485.30 \\ 
        fastText & 30.16 & 599.11 & 0.03 & 0.54 & 30.19 & 599.65 & 0.04 & 90.3 & 0.02 & 0.69 & 0.01 & 4.12 & 0.02 & 0.44 & 0.16 & 95.55 & 30.03 & 504.10 \\ 
        fzy & 2.70 & 54.07 & 0.02 & 0.36 & 2.72 & 54.44 & 0.57 & 14.05 & 0.02 & 0.57 & 0.06 & 6.76 & 0.01 & 0.12 & 0.72 & 21.50 & 2.00 & 32.94 \\ 
        gravity & 19.14 & 383.52 & 0.05 & 0.91 & 19.18 & 384.43 & 1.09 & 77.52 & 0.26 & 5.28 & 0.05 & 8.63 & 0.02 & 2.12 & 2.20 & 93.55 & 16.98 & 290.88 \\ 
        libco & 13.51 & 269.49 & 0.03 & 0.51 & 13.53 & 270 & 1.04 & 32.16 & 0.14 & 2.88 & 0.12 & 2.84 & 0.03 & 0.56 & 1.32 & 38.44 & 12.21 & 231.56 \\ 
        python & 312.94 & 6,233.32 & 1.78 & 35.57 & 314.72 & 6,268.88 & 15.20 & 669.16 & 0.53 & 22.54 & 0.01 & 15.07 & 0.44 & 9.42 & 16.46 & 716.19 & 298.26 & 5,552.69 \\ 
        redis & n$\backslash$a & n$\backslash$a & n$\backslash$a & n$\backslash$a & n$\backslash$a & n$\backslash$a & 106.95 & 2,360.85 & 0.54 & 13.29 & 2.80 & 57.77 & 0.14 & 3.10 & 110.43 & 2,435.01 & n$\backslash$a & n$\backslash$a \\ 
        OpenCV & n$\backslash$a & n$\backslash$a & n$\backslash$a & n$\backslash$a & n$\backslash$a & n$\backslash$a & 24.99 & 8,992.86 & 4.44 & 102.47 & 9.51 & 180.32 & 16.52 & 330.99 & 55.11 & 9,606.64 & n$\backslash$a & n$\backslash$a \\ 
        \bottomrule
    \end{tabular}
      \begin{tablenotes}
        \fontsize{6.5pt}{7.5pt}\selectfont
        \item IBGraph: actual build dependency graph of incremental builds.~~~~~~~~~~~~~~~~~~~~~~~~~~~~~~~~~~~~~~~~~~~~~~~~~~~~~~~Med: Median 
     \end{tablenotes}
     \end{threeparttable}
    \vspace*{-4.0ex}    
    \label{tab:time consumption}
\end{table*}

\paragraph{Results from 240 commits} With the benefits of incremental builds and inferring actual build dependencies, \textsc{EChecker} significantly outperforms Buildfs in terms of run-time performance. 
For example, Buildfs for \textit{python} takes about a median of 314.72s and a total of 6,268.88s, while \textsc{EChecker} takes only a median of 16.46s and a total of 716.19s. \textsc{EChecker} eliminates the need to perform multiple clean builds, thus significantly improving the efficiency of detecting dependency errors for large-scale projects. In addition, \textsc{EChecker} relies on incremental builds with the project that do not modify the program structure and are therefore more pervasive. Almost all projects benefit from the incremental build system since it is an advanced feature of a mature and stable build system.

\begin{figure}[!t]
  \centering
  \includegraphics[scale=0.18,trim=0 00 20 25,clip]{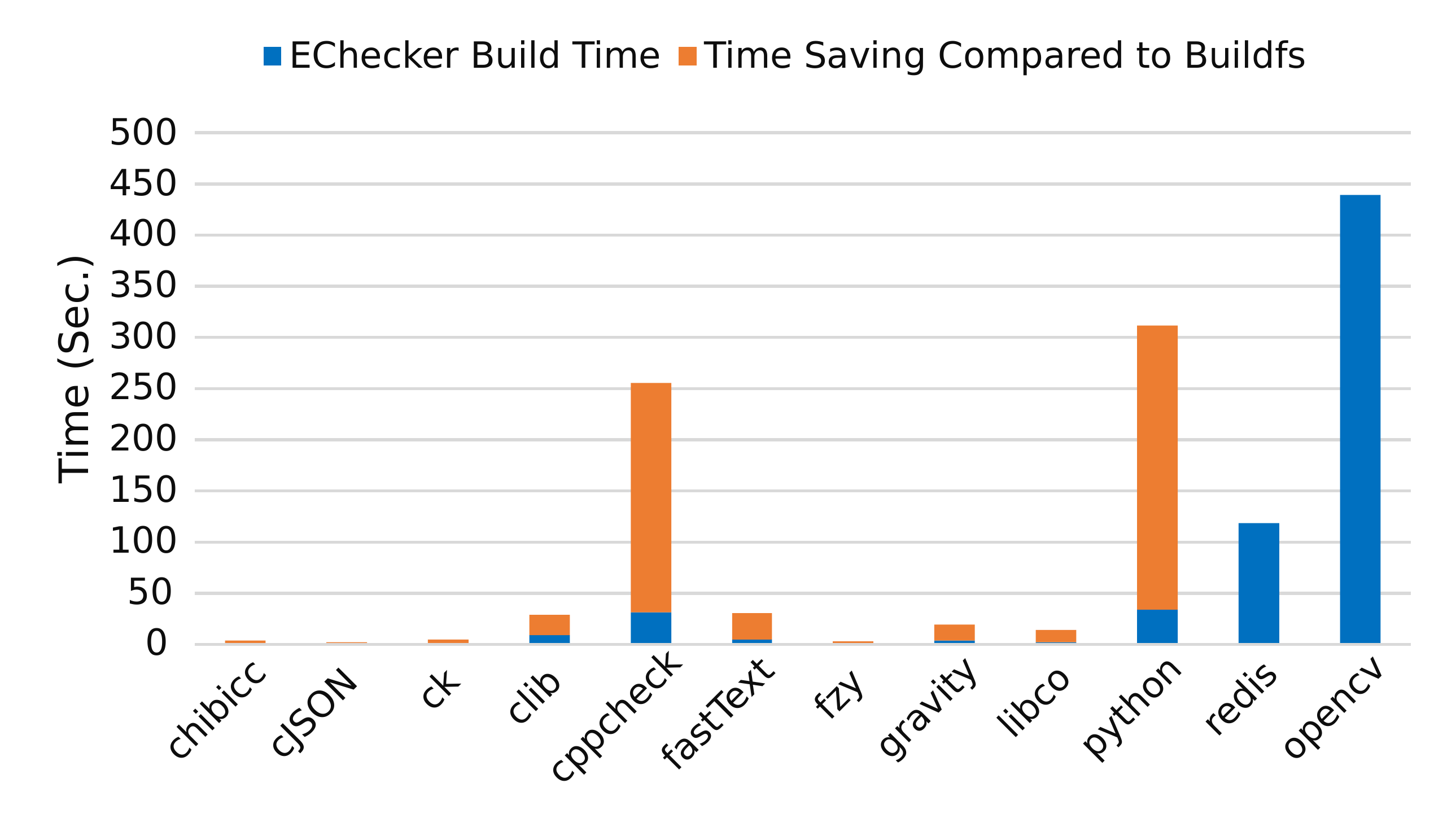}
  \caption{Average Time Saving Compared to Clean Builds}
  \label{FIG:Time Savings Compared to Clean Builds Using Incremental Builds.}
  \vspace{-2.0ex}
\end{figure}

\begin{figure}[!b]
  \vspace{-3.5ex}
  \centering
  \includegraphics[scale=0.55,trim=0 00 20 25,clip]{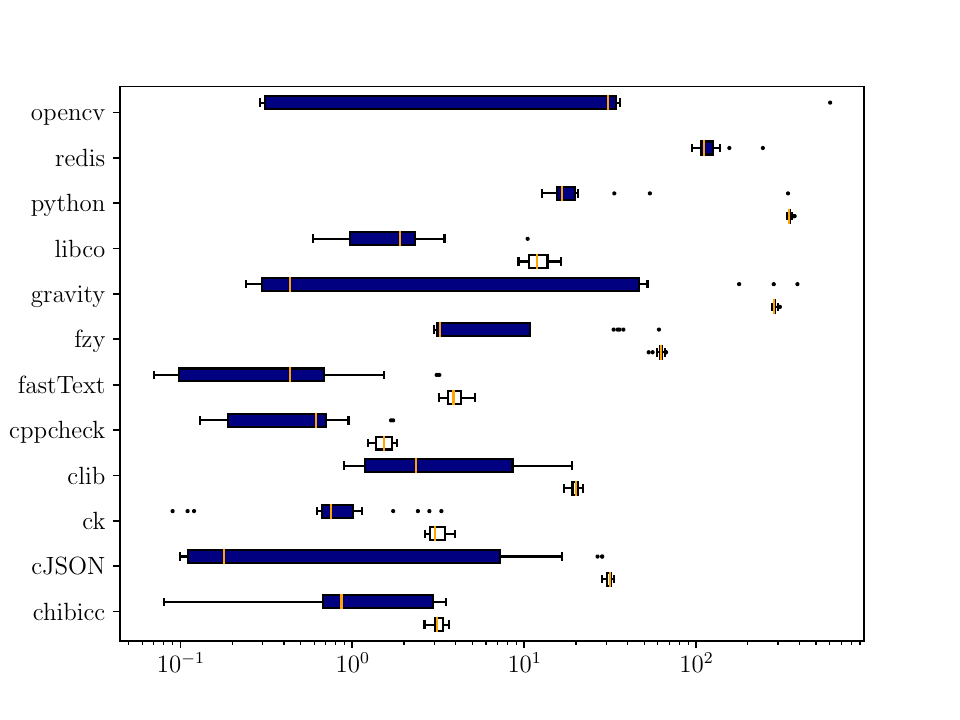}
  \caption{Sum of Time Consumption (Sec.) for Buildfs and \textsc{EChecker} to Detect in Each Commit. White Blox is Buildfs, Blue Blox is \textsc{EChecker}, and Orange Line is Median }
  \label{FIG:Sum of Time Consumption}
\end{figure}

\paragraph{Time savings for the build} To quantify the time saving for the build, we measure the time taken by \textsc{EChecker} and Buildfs to execute the build individually. To ensure fairness, each build is executed twice independently in a clean environment. Clean builds and incremental builds are executed independently for each project.
\Cref{FIG:Time Savings Compared to Clean Builds Using Incremental Builds.} shows the time saving of the 12 projects. 
The results show that \textsc{EChecker} using incremental builds can significantly reduce build time, in particular for large projects. For example, in the large-scale project (\textit{python}), incremental builds saved a median of 297.74s (312.94-15.2=297.74), which is 95.1\% (297.74s/312.94s) of the clean build time. In addition, in small-scale projects, incremental builds are still effective in saving build time. For example, in \textit{fzy}, incremental builds save a median of 2.13s, which is 78.9\% (2.13s/2.70s) of the clean build time. In \textit{cJSON}, using incremental builds reduces build time by a median of 75.6\% (1.3s/1.72s). 

\paragraph{Efficiency comparison} As shown in~\Cref{tab:time consumption} and \Cref{FIG:Sum of Time Consumption}, \textsc{EChecker} improves detection time by 85.14 times on average (with the median at 16.30 times) in comparison to Buildfs. In particular, the performance advantage of \textsc{EChecker} is pronounced on large-scale projects (more than 1,000 files), such as \textit{cppcheck} and \textit{python}. For \textit{cppcheck}, \textsc{EChecker} has an average improvement of 553.20 times (median is 659.07 times) over Buildfs, with a maximum improvement of 1,103.72 times (Buildfs 259.37s \textit{vs.} \textsc{EChecker} 0.24s). 
For \textit{redis} and \textit{OpenCV}, Buildfs failed to perform the detection, but the clean build time for these two projects on our machine is as high as 242.32s and 8,521.85s. \textsc{EChecker} only takes 110.43s and 55.11s on the median, which demonstrates its efficiency. In addition, \textsc{EChecker} also has a smaller median than Buildfs in terms of detection time consumption.
This is, because existing methods, including Buildfs, are limited by the need to use clean builds, which are time-consuming on large-scale projects. In contrast, \textsc{EChecker} supports the use of incremental builds, allowing for rapid feedback. As a result, \textsc{EChecker} completes checks in a short time. 
Incremental builds do not always finish quickly, in certain projects (\eg \textit{redis}) for instance, \textsc{EChecker} has a slightly lower efficiency. The overall execution time of incremental builds is affected by the build size required. The more artifacts an incremental build system builds, the longer it will take. In most cases, incremental builds consume less time than clean builds. Only in the worst case, does the time consumption of an incremental build equal that of a clean build.
Such cases are the exceptions that do not affect the efficiency of \textsc{EChecker}.
 
\begin{table}[!t]
    \centering
     \caption{Time Consumption on Detecting Build Dependency Errors in 68 Commits (Sec.)}
    \fontsize{7.1pt}{10pt}\selectfont
    \begin{threeparttable}
    \begin{tabular}{l|rr|rr|rr|r}
    \toprule

        \multirow{2}{*}{Projects} & \multicolumn{2}{c|}{Buildfs Sum} & \multicolumn{2}{c|}{Echecker Sum} & \multicolumn{2}{c|}{Difference Sum} & \multirow{2}{*}{Diff*} \\

    ~ & \multicolumn{1}{c}{Med} & \multicolumn{1}{c|}{All} & \multicolumn{1}{c}{Med} & \multicolumn{1}{c|}{All} & \multicolumn{1}{c}{Med} & \multicolumn{1}{c|}{All}& ~ \\ 
    \midrule
    chibicc & 3.46 & 37.76 & 0.71 & 17.25 & 2.75 & 20.51 & 0.06 \\
    cJSON & 1.73 & 18.75 & 0.3 & 6.52 & 1.43 & 12.23 & 1.29 \\
    ck & 4.92 & 38.72 & 3.68 & 29.34 & 1.24 & 9.38 & -3.14 \\
    clib & 29.14 & 197.46 & 2.72 & 75.19 & 26.42 & 122.27 & 0.58 \\
    cppcheck & 257.61 & 1,280.26 & 0.41 & 269.97 & 257.20 & 1,010.29 & 0.25 \\
    fastText & 30.11 & 149.42 & 0.54 & 30.97 & 29.57 & 118.45 & -0.46 \\
    fzy & 2.77 & 16.52 & 0.29 & 5.19 & 2.48 & 11.33 & 0.48 \\
    gravity & 19.15 & 384.43 & 2.86 & 34.14 & 16.29 & 350.29 & -0.69 \\
    libco & 13.75 & 69.39 & 1.49 & 18.32 & 12.26 & 51.07 & 0.05 \\
    python & 314.78 & 940.71 & 16.46 & 716.19 & 298.32 & 224.52 & 0.06 \\
    \bottomrule
    \end{tabular}
      \begin{tablenotes}
        \fontsize{6.5pt}{7.5pt}\selectfont
        \item *: \textsc{EChecker} at first baseline and second baseline, the difference in median sum detection time~~~~~~~~~~~~~~~~~~~~~~~~~~~~~~~~~~~~~~~~~Med: Median 
     \end{tablenotes}
     \end{threeparttable}
    \vspace*{-4.0ex}    
    \label{tab:time consumption in 68 commits}
\end{table}
\paragraph{Results from 68 commits}
As shown in~\Cref{tab:time consumption in 68 commits}, \textsc{EChecker} has a higher median total detection time in seven projects than the first baseline, ranging from 0.05s to 1.29s. The median total detection time decreased in three projects, ranging from 0.46 seconds to 3.14 seconds. Compared to Buildfs, \textsc{EChecker} increases the efficiency of error detection by 56.66 times (with a median of 5.93 times) in the second baseline. 

\vspace{1.0ex}
\begin{mdframed}[backgroundcolor=light-gray]
\small
\emph{\textbf{Answer to Q.2:} In the first baseline, \textsc{EChecker} detects dependency errors in 0.40s (\textit{ck})--55.11s (\textit{OpenCV}), an average improvement over Buildfs of 85.14 times (median is 16.30 times). In the second baseline, \textsc{EChecker} increases the build dependency error detection efficiency by an average of 56.66 times (with a median of 5.93 times). Overall,  \textsc{EChecker}'s performance decreased slightly in the second baseline compared to the first baseline, but the decrease is not significant, which demonstrates the efficiency of \textsc{EChecker}.}
\end{mdframed}

\section{Discussion}\label{6}
In this section, we further discuss the evaluation results, the generality, and the threats to the validity of this study.

\paragraph{Evaluation results}
To improve the credibility of our evaluation, we went through a process to confirm the false positives and negatives of errors reported by \textsc{EChecker}. We computed the ground truth of build dependency errors based on the naive build process and user declarations. We also used the Buildfs evaluation results as the baseline.
In practice, the false positive rate is an important indicator of the usefulness of the method. In total, \textsc{EChecker} detected 2,298 dependency errors, with only 32 false positives. These false positives are caused by improper build commands and can be quickly fixed by changing the build commands. There are two ways to quickly eliminate false positives when a practitioner discovers them: removing the artifact (\eg ``libcjson.so.1'') and re-detecting it or modifying the build commands (\eg ``ln -sf''). In comparison, Buildfs detected 1,600 errors, with 187 false positives (\ie phony targets) and 206 false negatives. The lower false positive rate might make \textsc{EChecker} a more practical alternative to Buildfs.

\textsc{EChecker} has better performance in terms of the time needed to detect build dependency errors (an average improvement over Buildfs of 85.14 times). This enables \textsc{EChecker} to efficiently and consistently assist practitioners in detecting build dependency errors in the current development model of continuous integration. In contrast, the state-of-the-art methods for detecting build-dependency errors (\eg Buildfs), which rely on clean builds, are difficult to use widely in practice due to the time consumption.

\paragraph{Generality}
We have designed \textsc{EChecker} for C/C++ and the widely-used build tools \textsc{GNU Make} and \textsc{CMake}. However, the core methodology used by \textsc{EChecker} is not only applicable to \textsc{GNU Make} and \textsc{CMake}, it is also applicable to other build systems such as \textsc{Gradle}~\cite{gradle23}, Ninja~\cite{Ninja12}, and Bazel~\cite{Bazel}. Adaptations to other build systems are required, for example, by modifying the implementation to monitor builds to adapt to \textsc{Gradle} builds. As \textsc[GNU Make] builds a process for each target, while Javac does not, it builds multiple targets in a single process~\cite{javac}. other build systems provide similar functionality to simulate builds. In addition, our approach implements the use of compiler directives to read build commands and build targets in \textsc{GNU Make}. Similar functionality is also provided by other build systems. For example, Gradle provides an API~\cite{GradlePlugins} to achieve declared inputs/outputs of every task and declared dependencies of every task (Gradle programmers assemble build logic in a set of tasks~\cite{Gradleuser}).

\paragraph{Threat to validity}
The main likely threat to the validity of our approach is the experimental environment of the machine, which may impact the time consumption of the experimental projects. Typically, the environment of the machine affects program execution time (\eg build time). This affects the validity of the efficiency evaluation. To minimize the random factor, we stopped other user processes in the system, repeated the evaluation twice for individual experimental projects, and chose the average time consumption of each stage as the final result. In addition, we calculated the standard deviation of the two experiment time consumption. The results show that 2 times experiments can effectively mitigate the impact on the validity of experimental results due to fluctuations in machine performance. Since the standard deviation ranges from 0.001s (0.03\% of sum detection time in \textit{cppcheck}) to 29.771s (0.03\% of sum detection time in \textit{OpenCV}, \cf supplementary materials).

The correctness of the way we have designed to calculate ground truth might be a threat. Previous studies~\cite{Bezemer17,Licker19,Sotiropoulos20,Fan20,Wu22} did not consider the ground truth; their approach validated detected errors. We computed the ground truth based on the real build process and declared build dependencies. The real build process provides the ground truth for MDs. Based on this, we removed the non-missing dependencies and then built each target individually to be able to compute the ground truth of RDs. This is because true RDs do not need to build the targets.

Another potential threat is that our results might not be generalizable to other projects. Additionally, the 20 commits for each project are also not a realistic frequency of code updates for practitioners in real development. We use 12 projects with 240 commits to evaluate the effectiveness and efficiency of our approach. While this project and commit number might appear small, all of these projects are active and well-known. All of these evaluated projects are mature and widely used, demonstrating that they may be representative of significant projects with build dependency errors.

\section{Related Work}
 We subsequently discuss the most relevant research related to detecting build dependency errors.

\paragraph{Detecting build dependency errors}
Due to the importance of build dependency errors~\cite{raymond2003art,Martin16,Beller17,Fitzgerald17}, the software engineering community has proposed many ways to detect errors. Research on detecting build dependency errors can be divided into two main categories based on whether or not it depends on the clean build.

The first category eschews clean build to detect build dependency errors. Such methods typically detect dependency errors using static analysis of source code and build scripts. Gunter et al.~\cite{Gunter96} used a Petri net to represent dependencies in build scripts to detect dependency errors in build scripts. Tamrawi et al.~\cite{Tamrawi12,TamrawiNNN12} proposed SYMake to automatically detect code smells in build scripts by constructing a symbolic dependency graph. Xia et al.~\cite{Xia14} statically analyzed build scripts and then used a link infection algorithm to infer missing dependencies. Zhou et al.~\cite{Zhou14} improved on Xia et al.'s method by using the declared build dependency graph and the source code. Both methods of Xia and Zhou were based on MAKAO~\cite{Adams07} and used only declared build dependencies to detect dependency errors. Their methods are efficient but ineffective because they lack actual build dependencies.

The second category relies on clean build to detect build dependency errors. Bezemer et al.~\cite{Bezemer17} used MAKAO to obtain a declared build dependency graph and then analyzed the build trace log to obtain the actual build dependencies, leveraging the two dependency graphs to detect dependencies. Licker et al.~\cite{Licker19} triggered incremental builds by modifying the timestamp of the source files and observed whether the target files had been rebuilt to detect dependency errors. However, their approach required multiple incremental builds for one detection, which was high time consumption. In addition, their approach of identifying errors by changes in the target timestamp introduced many false positives. Sotiropoulos et al.~\cite{Sotiropoulos20} treated each build target as a task, reasoning about the dependencies of each target separately, enabling it to support detecting build dependency errors in Java. Fan et al.~\cite{Fan20} used the actual and declared build dependency graphs to propose a unified dependency graph for detecting dependency errors. Wu et al.~\cite{Wu22} based on Fan et al.'s method proposed a virtual build per target instead of an actual build to speed up the detection of dependency errors. Their methods are efficient but lack generalizability because virtual builds require the removal of the program constructs, which can lead to build failures.
Practitioners using these clean build-dependent methods have difficulty being informed in a timely manner of build-dependent errors introduced during development.

\paragraph{Build systems and tools} Many build systems and tools have been proposed for detecting build dependency errors~\cite{Curtsinger18,Curtsinger22}. Rattle~\cite{Spall} automatically captured dependencies by monitoring the build process. Fabricate~\cite{fabricate} and Memoize~\cite{memoize} automatically filled in missing dependencies for build targets by monitoring the build process at run-time. Such monitoring-based systems will inevitably slow down the build, leading to a reduction in build time. Although these build systems tried to avoid build dependency errors, they were still in their infancy and not widely used by developers due to low build efficiency, immaturity of the technology, and high migration costs. Bazel~\cite{Bazel} creates an isolated environment for each build task and then avoids MDs by blocking the build process from accessing files that are not declared as dependencies. This avoided parallel build failures caused by missing dependencies and redundant dependencies. Other commercial tools, including IBM ClearCase~\cite{clearcase} and VESTA~\cite{vesta}, only checked the run-time state for dependency violations. There is a difficulty in detecting unimplemented dependencies using these methods.

In contrast to related studies, we propose a novel approach that uses code changes to infer actual build dependency graphs.
 
\paragraph{Build maintenance}
There is much work involved in the maintenance of software builds~\cite{Jafar14,Vakilian15,Jafar14B,Ren18,Ren19,Ren22}.
Gallaba et al.~\cite{Gallaba22} proposed a way to decouple the acceleration of the build from the underlying build tools, speeding up the build by inferring dependencies between build steps. Build failures are common bugs. It is also urgent for developers to locate and fix the cause of the build failure. Many methods are based on historical data~\cite{Hassan18}, searching~\cite{Lou19}, and deep learning frameworks~\cite{Tarlow20}.

\section{Conclusion}\label{8}
In this paper, we have proposed a new approach, called \textsc{EChecker}, to detect build dependency errors. Unlike the existing methods~\cite{Bezemer17,Licker19,Fan20,Sotiropoulos20,Wu22}, which rely on clean builds, \textsc{EChecker} avoids clean builds when possible to obtain the actual build dependency graph. \textsc{EChecker} uses pre-processor directives and Makefile changes to infer actual build dependencies. As a result, \textsc{EChecker} no longer requires multiple clean builds.
We selected 12 representative projects with 240 commits for detecting build dependency errors to evaluate the effectiveness and efficiency of \textsc{EChecker}. The evaluation results show that \textsc{EChecker} outperforms the state-of-the-art method (Buildfs) in terms of effectiveness and efficiency, as well as demonstrate that \textsc{EChecker} has a direct impact on detecting build dependency errors for practitioners in real-world applications, as it significantly improves detection efficiency. Therefore, practitioners will be able to identify and address build dependency errors during the localization process promptly.

\section*{Acknowledgments}
This work is supported by the National Natural Science Foundation of China (No.62202219, No.62072227, No.62302210), the Jiangsu Provincial Key Research and Development Program (No.BE2021002-2), and the Innovation Project and Overseas Open Project of State Key Laboratory for Novel Software Technology (Nanjing University) (ZZKT2022A25, KFKT2022A09, KFKT2023A09, KFKT2023A10). 

\balance
\bibliographystyle{ACM-Reference-Format}
\bibliography{main}
\end{document}